\documentclass[useAMS,usenatbib]{mn2e}
\usepackage{txfonts}
\usepackage{natbib}
\usepackage[all]{xy}
\usepackage[dvips]{graphicx}
\usepackage{hyperref}
\usepackage[usenames,dvipsnames]{color}
\usepackage{multirow}
\definecolor{menublue}{rgb}{0.0,0.0,0.5}
\definecolor{citegreen}{rgb}{0.0,1.0,0.0}
\definecolor{urlred}{rgb}{1.0,0.0,0.0}
\hypersetup{bookmarksopen,pdfstartview={FitH},colorlinks=true, breaklinks=true,menucolor=menublue,urlcolor=urlred,citecolor=citegreen,linkcolor=blue}
\usepackage[all]{hypcap}
\usepackage{epsfig}
\usepackage{epspdfconversion}
\usepackage{aas_macros}
\def\del#1{{}}

\newcommand{\ltsima}{$\; \buildrel < \over \sim \;$}
\newcommand{\lsim}{\lower.5ex\hbox{\ltsima}}
\newcommand{\gtsima}{$\; \buildrel > \over \sim \;$}
\newcommand{\gsim}{\lower.5ex\hbox{\gtsima}}

\newcommand{\dd}{\mathrm{d}}

\newcommand{\be}{\begin{equation}}
\newcommand{\ee}{\end{equation}}
\newcommand{\bea}{\begin{eqnarray}}
\newcommand{\eea}{\end{eqnarray}}
\newcommand{\dep}{\delta p}
\newcommand{\lkr}{{\left(k\right)}}
\def\cs{c_{\rm s}^2}
\def\ce{c_{\rm eff}^2}
\def\cv{c_{\rm v}^2}
\def\ca{c_{\rm a}^2}
\def\Om{\Omega_{m,0}}
\def\ceff{c_{\rm eff}^2}

\title[viscous dark energy]
{Combined constraints on deviations of dark energy from an ideal fluid from Euclid and Planck}
\author[E. Majerotto, D. Sapone, B.M. Sch{\"a}fer]
{Elisabetta Majerotto$^1$\thanks{e-mail: elisabetta.majerotto@uam.es}, Domenico Sapone$^2$ and Bj{\"o}rn Malte Sch\"afer$^3$\\
$^1$Instituto de F{\'i}sica Te{\'o}rica, Universidad Aut{\'o}noma Madrid, C/ Nicol‡s Cabrera 13-15, Cantoblanco, 28049 Madrid, Spain\\
$^2$Cosmology and Theoretical Astrophysics group, Departamento de F{\'i}sica, FCFM, Universidad de Chile, Blanco Encalada 2008, Santiago, Chile\\
$^3$Astronomisches Recheninstitut, Zentrum f{\"u}r Astronomie der Universit{\"a}t Heidelberg, Philosophenweg 12, 69120 Heidelberg, Germany}

\begin{document}
\pagerange{\pageref{firstpage}--\pageref{lastpage}}
\pubyear{2015}
\maketitle
\label{firstpage}

% --- abstract --- %
\begin{abstract}
Cosmological fluids are commonly assumed to be distributed in a spatially homogeneous way, while their internal properties are described by a perfect fluid. As such, they influence the Hubble-expansion through their respective densities and equation of state parameters. The subject of this paper is an investigation of the fluid-mechanical properties of a dark energy fluid, which is characterised by its sound speed and its viscosity apart from its equation of state.
In particular, we compute the predicted spectra for the integrated Sachs-Wolfe effect for our generalised fluid, and compare them with the corresponding predictions for weak gravitational lensing and galaxy clustering, which had been computed in previous work.  
We perform statistical forecasts and show that the integrated Sachs-Wolfe signal obtained by cross correlating Euclid galaxies with Planck temperatures, when joined to galaxy clustering and weak lensing observations, yields a percent sensitivity on the dark energy sound speed and viscosity. We prove that the iSW effect provides strong degeneracy breaking for low sound speeds and large differences between the sound speed and viscosity parameters.
\end{abstract}

% --- keywords --- %

% --- section: introduction --- %
\section{Introduction}
The expansion dynamics of the Universe is usually described by assuming $(i)$ general relativity as the theory of gravity, $(ii)$ a high degree of symmetry, namely spatial isotropy and homogeneity at each instant in time, and $(iii)$ ideal fluids which source the gravitational fields. These three assumptions lead to the Friedmann-equations for the time-evolution of the scale factor $a(t)$, which reflect the fact that Einstein's field equation is of second order, and shows acceleration or deceleration $\ddot{a}$ as phenomena.

The inclusion of the cosmological constant on grounds of the Lovelock theorem, which states that the field equation is the most general one in four dimensions, which includes derivatives of the metric of up to second order, and which conserves energy-momentum, yields a natural way to explain cosmic acceleration at late times.

Introducing dynamic dark energy components based on scalar self-interacting fields and interpreting the energy-momentum-tensor with the corresponding conservation law allows the identification of the homogeneous and isotropic field with a relativistic ideal fluid, whose relation between pressure and density is parametrized by an equation of state $w=p/\rho$. At the same time, this equation of state is the only free function that is allowed by the Einstein field equation with the symmetry assumptions of the Robertson-Walker metric.

For these reasons, a central goal of cosmology is to investigate dark energy and the cosmological constant through their influence on the dynamics of the scale factor and on the growth of structures. The fluid-picture is attractive due to its generality: Apart from actual substances like relativistic components ($w=+1/3$) and nonrelativistic components ($w=0$), it is general enough to describe spatial curvature ($w=-1/3$) and the cosmological constant ($w=-1$), while isotropy and homogeneity of the fluid ensure the Friedmann-symmetries.

Dark energy models based on self-interacting scalar fields show a natural variation of the dark energy equation of state parameter, because their time-evolution is governed by the Klein-Gordon equation, and therefore the kinetic and potential terms in their energy-momentum tensor evolve, leading to a time evolution in the equation of state, and therefore to a variation of their influence on the expansion dynamics of the Universe. In the slow-roll limit one recovers values of $w$ close to $-1$, resulting in accelerated expansion.

Adopting the fluid picture is an important test of whether the dark energy fluid is ideal or not: If the fluid has inhomogeneities, pressure fluctuations and density fluctuations are related through a sound speed, which in the most straightforward case describes an adiabatic compression of the fluid, there can be anisotropic stresses, and finally, velocity perturbations can experience viscous forces that dissipate kinetic energy \citep[for literature in this field we refer to][]{Battye:2006mb, Mota:2007sz, Battye:2009ze, Calabrese:2010uf, Ballesteros:2011cm, fingerprinting3, fingerprinting4, Appleby:2013upa, Dossett:2013npa, 2013JCAP...01..004S, Amendola:2013qna, Chang:2014mta, Chang:2014bea, Cardona:2014iba, Pearson:2014iaa, Ballesteros:2014sxa}. In addition, there can be a nonlinear relation between pressure and density of a fluid, one example of which would be Chaplygin-cosmologies \citep{Bento:2002ps, Li:2014bsa}, while a similar phenomenology could in principle be due to modifications in gravity rather than due to non-ideal fluids under general relativity \citep{Kunz:2006ca, Bertschinger:2008zb, Silvestri:2009jw, Pogosian:2010tj, 2010JCAP...04..018S, Leon:2010pu, Saltas:2010tt, 2013PhRvD..87b4015B, Boubekeur:2014uaa}.

These modifications break homogeneity on small scales, and require corresponding fluid equations for their time evolution, as well as couplings to local gravitational fields, which enable interaction between the dark energy fluid, the dark matter and the baryonic component. Commonly, one observes a difference between the two metric potentials in the case of nonzero sound speeds and equations of state unequal to $-1$, which can be probed by photons, relative to the motion of nonrelativistic objects such as galaxies, which is only sensitive to a single metric potential.

In this paper we investigate cosmological perturbations with a non-ideal dark energy fluid and aim to forecast constraints on its speed of sound $c_s$ and its viscosity from Euclid\footnote{http://www.euclid-ec.org/}  \citep{Cimatti:2009is, RedBook} and Planck \citep{planck2013, planck2015}. Specifically, we consider tomographic weak gravitational lensing \citep{2012MNRAS.422.3056A}, galaxy clustering \citep{2003PhRvD..67j3509D, 2006PhRvD..74d3505T} and the integrated Sachs-Wolfe effect \citep{Dossett:2013npa, Soergel:2014sna} as probes on the influence of non-ideal fluids on the statistics and the evolution of structures, while the background expansion dynamics is given through the individual density parameters and the equation of state parameters, assuming that there is no energy exchange between the fluids.

Our work is complementary to that of \citet{Mota:2007sz, Calabrese:2010uf, Chang:2014mta}, who used the same model to describe the evolution of anisotropic stress: \citet{Mota:2007sz, Chang:2014mta} computed constraints on it from the Cosmic Microwave Background (CMB), the large scale structure and Supernovae Type Ia, while \citet{Calabrese:2010uf} forecasted errors from the CMB on the parameters of an early dark energy possessing anisotropic stress. It is also complementary to that of \citet{Amendola:2013qna, Cardona:2014iba, 2013JCAP...01..004S}, who also put constraints on anisotropic dark energy, but used different models for its evolution.

Currently, there are no significant deviations from dark energy being a perfect fluid, for instance the result by \citep{2004PhRvD..69h3503B} who find $c_\mathrm{s}^2<0.04$ at low significance from CMB-data, such that tests whether dark energy is an ideal fluid will be the domain of future experiments: Quite generally, the sensitivity to non-ideal cosmic fluids requires their respective density to be large enough and their equation of state not to be too negative for dark energy perturbations to be sufficiently strong \citep{2002PhRvL..88l1301E, Koivisto:2005mm,  2010PhRvD..81j3513D, 2010JCAP...10..014B, Archidiacono:2014msa}. At a first sight it would appear that choosing a dark energy equation of state too far from the cosmological constant value is incompatible with present constraints \citep{planck2015}. However, when including extra parameters such as the speed of sound and viscosities in the fluid, constraints become much more loose \citep[See e.g.]{Mota:2007sz, Archidiacono:2014msa}.

This article is structured as follows: We develop the necessary perturbation equations for non-ideal dark energy fluids and a suitable parametrization in Sect.~\ref{sect_fluids} and discuss cosmological probes in Sect.~\ref{sect_probes}, before computing forecasts on non-ideal dark energy properties in Sect.~\ref{sect_forecasts}. We summarise our results in Sect.~\ref{sect_summary}. The reference cosmological model is a spatially flat, dark-energy dominated model with the parameter choices $\{\Omega_m h^2,\,\Omega_b h^2,\,n_s,\, \Omega_m \} = \{ 0.142,\, 0.022,\,0.67,\,0.96,\,0.32\}$. This corresponds to the constraints from Planck \citep{planck2013} and WMAP polarization  low-multipole  likelihood \citep{Bennett:2012zja, Ade:2013kta}, and it represents the present official baseline for Euclid forecasts. The amplitude of the primordial power spectrum was fixed to $A_s =  2.1 \times 10^{-9}$. The dark energy equation of state parameter was set to be constant with a numerical value of $w=-0.8$.

% ---  --- %
\section{cosmology with non-ideal fluids}\label{sect_fluids}

\subsection{Expansion dynamics}
Since we focus on late cosmological times, where dark matter and dark energy are dominating the energy density of the Universe, we can approximate the Hubble function $H(a)=\dot{a}/a$ with 
\be
H^2 = H_0^2\left[ \Omega_{m,0} a^{-3}+ (1-\Omega_{m,0} )a^{-3(1+w)} \right] ,
\ee
where $a$ is the scale factor, $\Omega_{m,0}$ is the dark matter density parameter today, $w$ is the equation of state of dark energy, which we assume to be constant, and $H_0$ is the Hubble parameter today. In addition, we do not consider global curvature. The comoving distance is defined as
\be \label{eq:comdist}
\chi = c \int_a^1  \frac{da}{a^2 H(a)} \equiv \chi_H \int _a^1 \frac{da}{a^2H(a)/H_0},
\ee
with the Hubble distance $\chi_H=c/H_0\simeq2996.9~\mathrm{Mpc}/h$. At the same time, this defines conformal time $\tau$ through $\chi=c\tau$. In the following, we will set $c=1$.

% ---  --- %
\subsection{Perturbations and their analytical solutions} \label{subsect_pert}

If we consider a non-ideal fluid dark energy, characterised by a constant equation of state $w$, a speed of sound $c_s$, and an anisotropic stress component $\sigma$, we can write the evolution of $\sigma$ as in \cite{Hu:1998kj}:
\be
\sigma' + \frac{3}{a} \sigma = \frac{8}{3} \frac{\cv}{(1+w)^2} \frac{V}{a^2H}, \label{eq:sig}
\ee
where the prime indicates derivative with respect to $a$ and $\cv$ is called viscosity parameter, as it gives a measure of the fluid's viscosity: Indeed, Eq.~(\ref{eq:sig}) implies that when $\cv = 0$, then the anisotropic stress component $\sigma$ is also vanishing, while when e.g. $\cv = 1/3$ the evolution of anisotropic stress for radiation up to the quadrupole is recovered.

To this equation, we add the first order perturbation equations for the density contrast $\delta$ and the velocity perturbation $V$
\bea
\delta' &=& 3(1+w) \phi' - \frac{V}{Ha^2} - 3 \frac{1}{a}\left(\frac{\dep}{\rho}-w \delta \right) \label{delta} , \\
V' &=& -(1-3w) \frac{V}{a}+ \frac{k^2}{H a^2} \frac{\dep}{\rho}+(1+w) \frac{k^2}{Ha^2} \psi +\\ \nonumber
&-&(1+w)\frac{k^2}{Ha^2}\sigma , \label{v}
\eea
where $\dep$ is the 
pressure perturbation, $\rho$ is the dark energy density, $\psi$ and $\phi$ are the metric perturbations in the Newtonian gauge, defined by the line element 
\be
\dd s^2 = a^2 \left[ -(1+2\psi) \dd\tau^2 + (1-2\phi) \dd x_i \dd x^i \right]. \label{metric}
\ee
Pressure perturbations are parametrized as 
\be
\delta p = \cs \rho \delta + \frac{3 a H(\cs - \ca)}{k^2} \rho V ,
\ee
where $\ca \equiv \dot{p}/\dot{\rho} = w$ is the adiabatic speed of sound  for a fluid with constant equation of state, to which $\cs$ reduces in the case of a perfect fluid, when  no dissipative effects, leading to entropic perturbations, are present \citep{2004PhRvD..69h3503B}.

In order to close the differential equation system, one needs to include the Poisson equation
\be
k^2\phi = -4\pi G a^2\sum_i\rho_i\left( \delta_i+\frac{3aH}{k^2}V_i\right) = -4\pi G a^2\sum_i\rho_i\Delta_i \,, \label{eq.Poisson_general}
\ee
(where the sum runs over all clustering fluids, $G$ is the Newton constant, and in the last equality we have defined the gauge-invariant density perturbation of the $i$-th fluid, $\Delta_i \equiv \delta_i + 3 a H V_i/k^2$)
and the fourth Einstein equation
\bea  \label{eq.4thEinsteinB}
k^2\left(\phi -\psi \right) &=& 12 \pi G a^2 \, (1+w) \rho\, \sigma  \\
&=& \frac{9}{2}  H_0^2 (1-\Om)a^{-(1+3 w)} (1+w)  \sigma  \nonumber \\
&\equiv& B(a)\,\sigma \,. \label{eq.4thEinstein}
\eea
In \cite{fingerprinting3} the following analytical solutions for $\delta$, $V$ and $\sigma$ were found for the matter dominated era:
\bea
\delta &=& \frac{3(1+w)^2}{3 \cs (1+w) + 8 \left( \cs -w \right) \cv} \frac{\phi_0}{k^2}  ,  \label{eq:delta-sub-below}\\
V &=& - \frac{9(1+w)^2 \left( \cs -w \right)}{3 \cs (1+w) + 8 \cv (\cs -w)} H_0 \sqrt{\Omega_m} \frac{\phi_0}{\sqrt{a}k^2} , \nonumber \\
&=& -3  a H \left(\cs -w \right) \delta , \label{eq:V-sub-below}\\
\sigma &=& - \frac{8 \cv \left( \cs -w\right)}{3 \cs(1+w) + 8 (\cs -w) \cv} \frac{\phi_0}{k^2} ,
\label{eq:sigma-sub}
\eea 
where $k^2 \phi \simeq -\phi_0$, 
which is valid strictly only during matter domination and while neglecting dark energy perturbations.

As found in \cite{fingerprinting3}, to which we refer for further detail on the analytic solutions, the relevant quantity is the effective sound speed
\be
\ce= \cs+\frac{8}{3}\cv\frac{\cs-w}{1+w}\,,
\label{eq:ceff}
\ee
as Eqs. (\ref{eq:delta-sub-below}-\ref{eq:sigma-sub}) can be rewritten in terms of it. This means that the sound speed and the viscosity have a similar damping effect on density and velocity perturbations \citep[as also noticed in][]{Mota:2007sz, Calabrese:2010uf}. It is interesting to notice that the effect of $\cv$ is enhanced with respect to that of $\cs$ by a factor of $8(\cs-w)/[3(1+w)]$, which is $\sim 10$ if $w \sim -0.8$ and for very small $\cs$, which are the cases where a viscosity can be observed best, as will be shown in the following sections, and bounded by the case of a cosmological constant, as Eq.~(\ref{eq:delta-sub-below}) diverges for $w=-1$.

% ---  --- %
\subsection{Observable parameters}
To understand how the viscosity affects the physical observables, it is useful to introduce the clustering parameter $Q$  and the anisotropy parameter $\eta$, defined in \cite{Amendola:2007rr} and computed in the case of viscous dark energy in \cite{fingerprinting3}.

$Q$  parametrizes the 
deviation from a purely matter-dominated Newtonian potential and is given by \citep[see][]{fingerprinting3}
\bea
Q -1  &\equiv& \frac{\rho\Delta}{\rho_m\Delta_m} = \frac{1-\Om}{\Om}(1+w)\frac{a^{-3w}}{1-3w + \frac{2 k^2 a}{3H_0^2\Om}\ceff} \nonumber \\
&=& Q_0\frac{a^{-3w}}{1+\alpha\,a},
\label{eq:qtot}
\eea
where $\alpha =  2 k^2 \ce/[(3 H_0^2 \Om)(1-3w)]$ and $Q_0 = (1+w)(1-\Om)/\left[\Om (1-3w)\right]$.

The anisotropy parameter is then given by
\be \label{eq:eta-def}
\eta \equiv \frac{\psi}{\phi} -1 = -\frac{9}{2}H_0^2(1-\Om)(1+w)\frac{a^{-1-3w}}{k^2 Q}\left(1- \frac{\cs}{\ceff} \right)\,\label{eq:etatot}.
\ee
This is nonzero only when anisotropic stress is present and the metric perturbations $\phi$ and $\psi$ are different.

Let us finally define the parameter $\Sigma$ \citep{Amendola:2007rr} as
\be
\Sigma = \left(1+\frac{1}{2} \eta \right) Q.
\ee
This is useful because it represents the deviation of the weak lensing potential $\Phi = \psi + \phi$ from its behaviour in the case of no dark energy perturbations.

% --- section: cosmological probes ---%
\section{cosmological probes}\label{sect_probes}
In \cite{fingerprinting4}, we forecasted constraints to the viscosity parameter and the sound speed from the Euclid galaxy clustering and weak lensing surveys. Here, we aim to complete the picture by adding to the latter the constraints from the iSW tomography signal obtained by cross-correlating galaxies mapped by the Euclid photometric instrument with the Planck temperature map: This provides a combination of all major probes of cosmic structure formation, which draw their sensitivity from the growth rate and interactions between fluids, from the shape of the initial perturbations and from the expansion history.

The Euclid survey is a mission of the ESA Cosmic Vision program that will be launched in 2020, and will perform both a photometric and a spectroscopic survey, the first aiming mainly at measuring weak lensing while the second at measuring the galaxy power spectrum. The Planck satellite is also a mission of ESA Cosmic Vision program, already operating and mapping the CMB fluctuations with unprecedented precision and control of systematic effects.

To perform our forecasts we use the Fisher matrix \citep{Tegmark:1996bz}, which quantifies the decrease  in likelihood  if a model parameter $\theta_\alpha$ moves away from the fiducial value, and can be computed  for  a local  Gaussian approximation  to the likelihood $\mathcal{L} \propto \exp(-\chi^2/2)$.
In our forecasts we assume the official Euclid specifications, that can be found in  \cite{RedBook}.
The fiducial cosmological parameters correspond to the 2013 Planck measurements \citep{planck2013}, except for the value of $w$, for which we assume $w = -0.8$, in order for the effects to be more clearly visible \citep[as done in][]{fingerprinting3, fingerprinting4} and of course except for the values of $\cs$ and $\cv$.

In the following, we will describe the iSW tomography signal and give a short summary on the signal coming from weak lensing and the galaxy power spectrum from spectroscopy.

% ---  --- %
\subsection{iSW signal} 
When a CMB photon moves into a time-evolving metric such as that of Eq. (\ref{metric}) the unbalance between the blue-shift experienced at the entrance and the red-shift experienced at the exit of its varying  potential well originates a perturbation $\zeta$ in the CMB temperature $T_\mathrm{CMB}$  given by \citep{iSW},
\be \label{isw}
\zeta = 
\frac{\Delta T}{T_\mathrm{CMB}} \equiv 
\int\dd\tau\:\left( \frac{\partial \phi}{\partial \tau} + \frac{\partial \psi}{\partial \tau} \right) = \int_0^{\chi_H}\dd \chi\: a^2 H \frac{\partial \Phi}{\partial a},
\ee
where $\chi$ is the comoving distance (see Eq.~\ref{eq:comdist}) and $\Phi$ is the weak lensing potential.

In the case of pure matter domination, $\Phi = \mathrm{const}$, hence the iSW effect vanishes, while in presence of any fluid with $w \neq 0$ the temperature fluctuation $\zeta$ will be nonzero, so that the late iSW is particularly interesting to us, as it is originated by the appearing of dark energy and it is an independent proof of its existence \citep[first detected by][for an updated measurement]{Boughn:2003yz,Giannantonio:2012aa}. 

Let us now compute the term inside the integral, passing to Fourier space, and in the case of viscous dark energy \citep[see also][]{fingerprinting3, Schaefer:2008tz}:
\bea
&&\frac{\partial\Phi}{\partial a}=-\frac{3}{2} \frac{H_{0}^{2}\Om}{ak^2} \left\{\vphantom{\frac{1}{a}}\Sigma\left(a,k\right)\Delta'_{m}\left(a,k\right) + \right. \nonumber \\
&&+\left. \Sigma'\left(a,k\right)\Delta_{m}\left(a,k\right)-\frac{1}{a}\Sigma\left(a,k\right)\Delta_{m}\left(a,k\right)\right\}. 
\label{psidot}
\eea
It is possible to see from this expression that anisotropic  perturbations enter the iSW effect in two ways: by 
modifying $\Delta_m$ and through the additional presence of $\Sigma$ and $\Sigma'$.
At linear order, it is possible to isolate today's $\Delta_m$ from its time evolution: 
\be
\Delta_{m}\left(a,k\right) = aG\left(a,k\right)\Delta_{m,0}\left(k\right),
\ee
where $\Delta_{m,0}\left(k\right)\equiv \Delta_m(a=1,k)$.
We write hence Eq.~(\ref{psidot}) as:
\be
\frac{\partial\Phi}{\partial a} = -\frac{3}{2}\frac{H_{0}^{2}\Om}{k^2}\frac{\partial}{\partial a}\Big \{G\left(a,k\right) \Sigma\left(a,k\right) \Big\}\Delta_{m,0}\left(k\right)\,, 
\label{eq:psiprimeG}
\ee
and Eq. (\ref{isw}) reads now
\be
\zeta = \int_{0}^{\chi_{_H}}{\dd\chi\: W_{\zeta}\left(\chi\right)\Delta_{m,0}\lkr} 
\ee
where the weighting function $W_{\zeta} (\chi)$ is
\be
W_{\zeta}\left(\chi\right) =  \frac{3}{2}\frac{H_0^2\Om}{k^{2}}a^{2}H\frac{\partial}{\partial a}\Big \{G\left(a,k\right)\Sigma\left(a,k\right) \Big\}.
\label{windowISW}
\ee
Since the iSW is a secondary effect of the CMB \citep{Rees:1968zza}, it can be separated through cross-correlation to the galaxy density \citep{Crittenden:1995ak}. Let us hence write the galaxy density obtained through imaging surveys, in order to compute its cross-correlation with the iSW.
The line-of-sight projected galaxy density $\gamma$ is given by \citep{Smail:1995us}
\be \label{eq:gamma}
\gamma = \int_{0}^{\chi_H}{{\rm d}\chi\,D(z)\frac{{\rm d}z}{{\rm d}\chi}b(\chi)G(\chi)\delta(z)}
\ee
being $D(z)$ the galaxy distribution defined as
\be \label{eq:gal-distr}
D(z) = \left(\frac{z}{z_0}\right)^2\exp\left[-\left(\frac{z}{z_0}\right)^{\beta_D}\right],
\ee
where the fiducial parameters $\beta_D$ and $z_0$ depend on the imaging survey considered. In the case of Euclid, they are $\beta_D = 3/2$, $z_0=z_\mathrm{mean}/\sqrt{2}$, and $z_\mathrm{mean} = 0.9$ \citep{RedBook}.

Even though the signal from the iSW increases noticeably when cross-correlating it with the galaxy density field, both the cross-correlation spectrum and the galaxy spectrum are line-of-sight integrated quantities, hence much information may be lost. For this reason we decide to use iSW tomography \citep{2008PhRvD..78d3519H, 2008A&A...485..395D, isw-tomo}, and in particular we divide the whole galaxy sample into 5 bins with equal number of galaxies (in order to match with the binning used by official Euclid documents for weak lensing tomography). To do this, we replace the galaxy distribution function $D(z)$, Eq. (\ref{eq:gal-distr}) in $\gamma$, Eq. (\ref{eq:gamma}), with the radial distribution function of galaxies in the $i$-th bin  $D_i(z)$, obtained by binning the overall distribution $D(z)$ and convolving it with the photometric redshift distribution function \citep{Amendola:2007rr}.

\begin{figure}
\resizebox{\hsize}{!}{\includegraphics{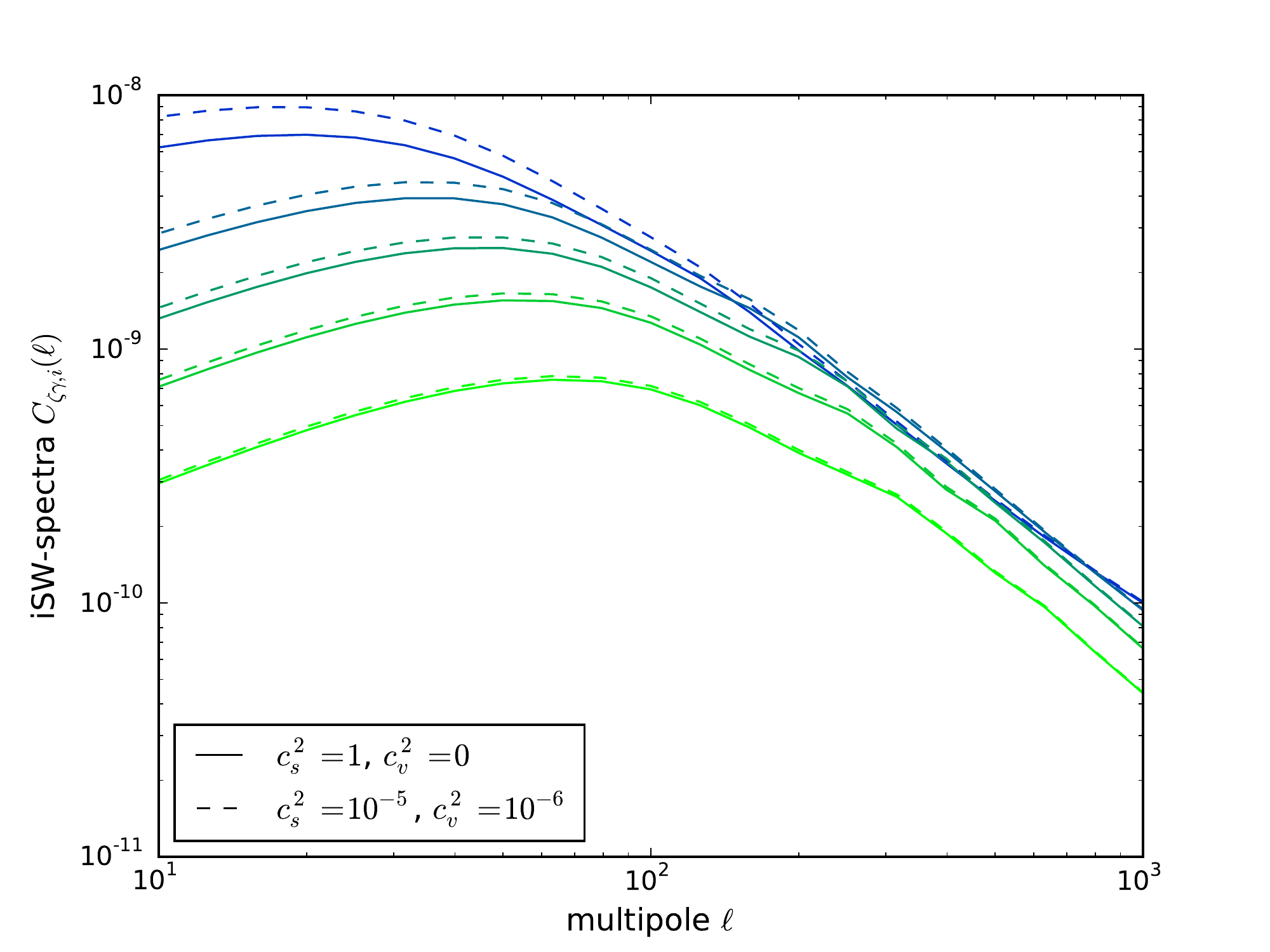}}
\caption{Tomographic iSW-spectra $C_{\zeta\gamma,i}(\ell)$ for two dark energy models: $\cs=1$ and $\cv=0$ as well as $\cs=10^{-5}$, $\cv=10^{-6}$. Blue to light green lines correspond to redshifts $z$ in the intervals $[0.01-0.5595]$ (blue), $[0.5595-0.7871]$, $[0.7871-1.0165]$, $[1.0165-1.3184]$ and finally $[1.3184-2.5]$ (green).}
\label{fig:isw-spectra}
\end{figure}

We are finally able to write our observable, i.e. the iSW-galaxy cross-correlation spectrum $C_{\zeta\gamma,i}(\ell)$ in the $i$-th redshift bin, along with the iSW-auto correlation spectrum $C_{\zeta\zeta}(\ell)$ and the galaxy-galaxy auto correlation spectrum $C_{\gamma\gamma,ij}(\ell)$ of the $ij$-bins, (which are both needed in order to estimate statistical errors coming from $C_{\zeta\zeta}(\ell)$), by applying a Limber-projection \citep{limber} in the flat-sky approximation:
\bea \label{cls-zetagamma}
C_{\zeta\gamma,i}(\ell) & = & 
\int_{0}^{\chi_H}\frac{{\dd\chi}}{\chi^2}\:W_{\zeta}(\chi)W_{\gamma,i}(\chi)\:P_{\Delta\Delta}(k=\ell/\chi),\\
\label{cls-zetazeta}
C_{\zeta\zeta}\left(\ell\right) & = & 
\int_{0}^{\chi_{_H}}\frac{\dd\chi}{\chi^2}\:W_{\zeta}^2 \left(\chi\right)\:P_{\Delta\Delta}\left(k =\ell/\chi\right)\\
C_{\gamma\gamma,ij}(\ell) & = & 
\int_{0}^{\chi_H}\frac{{\dd\chi}}{\chi^2}\:W_{\gamma,i}(\chi) W_{\gamma,j}(\chi)\:P_{\Delta\Delta}(k=\ell/\chi)
\label{cls-gammagamma}
\label{cls}
\eea
where $\bar{P}_{\Delta\Delta}\left(k\right)$ is the linear matter power spectrum today, and the galaxy weighting function of the $i$-th bin $W_{\gamma,i}(\chi)$ is
\be
W_{\gamma,i}(\chi) = D_i(z)\frac{{\rm d}z}{{\rm d}\chi}b(\chi)G(\chi).
\ee
The tomographic iSW spectra $C_{\zeta\gamma,i}(\ell)$ are shown in Fig.~\ref{fig:isw-spectra} for two fiducial models: a standard dark energy model with $\cs = 1$ and $\cv = 0$ (solid lines) and a model with viscosity: $\cs = 10^{-5}$ and $\cv = 10^{-6}$ (dashed lines), where the colour shading indicates the reshift bin for which $C_{\zeta\gamma,i}(\ell)$ was evaluated. The iSW-effect is a large-scale effect originating from low redshift, as the influence of dark energy on the growth of gravitational potentials in the large-scale structure is strongest. The effect of dark energy visosity and small sound speed is strongest on large scales as well \citep[see also][]{fingerprinting3}, and affects a wide range of multipoles. Keeping all cosmological parameters fixed, dark energy viscosity would increase the amplitude of the iSW-effect by up to 25\% on large angular scales and at low redshift. This sensitivity of the spectra at low multipoles is fortunate because these scales can be well probed with the iSW-effect.

% ---  --- %
\subsection{Weak Lensing} \label{subsec:wlps}
To the iSW tomography signal we add the weak lensing tomographic signal \citep{wl-tomo, wl-tomo-2, wl-tomo-3, wl-tomo-4}, coming  from the same photometric survey as $\gamma$ and using the same redshift bins. Here, we only give the main equation expressing the weak lensing power spectrum, which is used for our forecasts, and refer to \cite{fingerprinting4} for further details.

In presence of anisotropic stress, the weak lensing convergence power spectrum is given by \citep{wl-tomo-4,wl-tomo-2,wl-tomo,wl-fisher}
\begin{equation}
C_{\kappa,ij}(\ell) = 
\int_0^{\chi_H}\frac{\dd\chi}{\chi^2}\:W_{\kappa,i}(\chi)W_{\kappa,j}(\chi)\: \Sigma^2\: P_\mathrm{NL}(k=\ell/\chi,\chi).
\label{eq:convergence-wl}
\end{equation}
where the subscript ${ij}$ refers to the redshift bins around $z_i$ and $z_j$, with
\bea
W_{\kappa,i}(\chi) & = & \frac{3\Omega_m}{2\chi_H^2}\frac{F_i(\chi)}{a}\chi\\
F_i(\chi) & = & \int_\chi^{\chi_H}\dd\chi^\prime n(\chi^\prime)\:D_i(\chi^\prime)\:\frac{\chi^\prime-\chi}{\chi^\prime}
\eea
and where $D_i$ is the same tomographic distribution function of galaxies used for the iSW-effect. While tomography in general greatly reduces statistical errors the actual shape of the choice of the binning does not affect results in a serious way, although in principle there is room for optimisation \citep{2012MNRAS.423.3445S}.

\begin{figure}
\resizebox{\hsize}{!}{\includegraphics{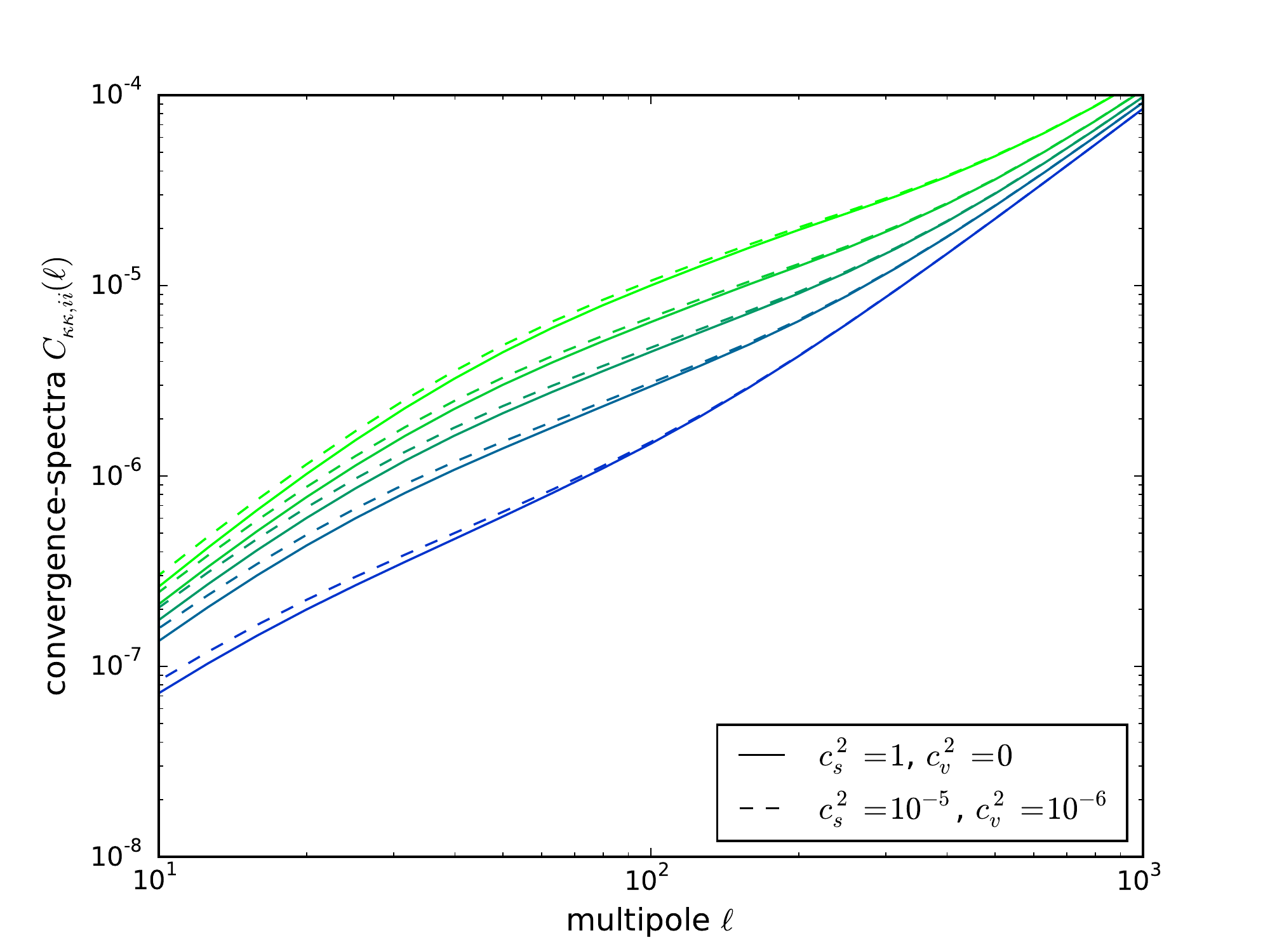}}
\caption{Tomographic weak lensing spectra $C_{\kappa\kappa,ii}(\ell)$ for two dark energy models, $c_s=1$ and $c_v=1$ as well as $c_s=10^{-5}$, $c_v=10^{-6}$, both including the shape noise term. Blue to light green lines correspond to the same redshift binning as in Fig.~\ref{fig:isw-spectra}: $[0.01-0.5595]$ (blue), $[0.5595-0.7871]$, $[0.7871-1.0165]$, $[1.0165-1.3184]$ and finally $[1.3184-2.5]$ (green).}
\label{fig:wl-spectra}
\end{figure}

In Fig.~\ref{fig:wl-spectra} we show the tomographic weak lensing spectra $C_{\kappa,ii}(\ell)$ for the same models and the same redshifts as in Fig.~\ref{fig:isw-spectra}. As for iSW, the effect of viscosity is detected at large scales and for a large range of scales (but smaller than for iSW). Instead, contrarily to iSW, here the sensitivity to viscosity is stronger at higher redshift. This is because the efficiency of weak lensing is higher for longer light paths.

In principle it is also possible to define a cross-spectrum of weak lensing and iSW, $C_{\kappa\zeta,i}(\ell)$, and of weak lensing and galaxy distribution, $C_{\kappa\gamma,i}(\ell)$, but both these spectra are subdominant with respect to $C_{\zeta\gamma}(\ell)$. This is because the weak lensing convergence signal comes from the distortion of the light path at redshifts intermediate between us and the galaxies mapped by the imaging survey, while the iSW signal originates precisely at the same redshifts where the galaxies are. We have tested this fact by computing the signal to noise-ratio for measuring $C_{\zeta\kappa,i}(\ell)$ and found it much smaller than that of $C_{\zeta\gamma,i}(\ell)$.

% ---  --- %
\subsection{Spectroscopic galaxy power spectrum}
To the iSW and weak lensing measurements, both measured through photometric observations, we add data coming from the power spectrum of spectroscopically observed galaxies. Here we only show the expression of the observed power spectrum, which is needed in order to compute our forecasts, and refer again the reader to \cite{fingerprinting4} for further detail.

Following \cite{seisen} we write the observed galaxy power spectrum as: 
\be
P_{\gamma\gamma}^\mathrm{spec}(z,k_r,\mu_r) =
\frac{D_{Ar}^{2}(z)H(z)}{D_{A}^{2}(z)H_{r}(z)} G^{2}(z,k)b(z)^{2}\left(1+\beta\mu^{2}\right)^{2}P_{0r}(k) +  P_\mathrm{shot} \,, \label{eq:pk}
\ee
where the subscript $r$ refers to the reference (or fiducial) cosmological model.

Here $P_\mathrm{shot}$ is a scale-independent offset due to imperfect removal of shot-noise, $\mu = \vec{k}\cdot\hat{r}/k$, is the cosine of the angle of the wave mode with respect to the line of sight pointing into the direction $\hat{r}$, $P_{0r}$ is the fiducial matter power spectrum evaluated at  redshift zero, $G (z,k)$ is the linear growth factor of the matter perturbations, $b(z)$ is the bias factor and $D_{A}(z)$ is the angular diameter distance. The wavenumber $k$ and $\mu$ have also to be written in terms of the fiducial
cosmology \citep[see][for more details]{seisen, aqg, sa}. The fiducial bias used can be found in \cite{Orsi:2009mj}, who derived their results by using a semi-analytical model of galaxy formation, while the matter power spectrum has been computed with a modified version of the CAMB code\footnote{http://camb.info} \citep{camb} accounting for anisotropies.

% --- section: forecasts ---%
\section{Statistical errors forecasts}\label{sect_forecasts}
In this section we estimate marginalised statistical errors on the sound speed and viscosity parameters $\cs$ and $\cv$ through the Fisher-matrix formalism \citep{Tegmark:1996bz}, which assumes a Gaussian likelihood and unbiased measurements.

% ---  --- %
\subsection{iSW Fisher matrix}
The sensitivity of line of sight-integrating effects can be boosted by subdividing the galaxy population into redshift bins: For the iSW-effect this was first carried out successfully by \citet{2008PhRvD..78d3519H}, and systematically investigated by \citet{isw-tomo}.

The Fisher-matrix of the iSW-effect follows directly from the variance of the spectrum estimates,
\be
F^{\rm iSW}_{\alpha\beta} = 
\sum_{\ell} \frac{\partial \bar{C}_{\zeta\gamma,i}(\ell)}{\partial \theta_\alpha}{\rm Cov}_{ij}^{-1}(\ell)\frac{\partial \bar{C}_{\zeta\gamma,j}(\ell)}{\partial \theta_\beta}
\ee
where the sum runs from $\ell = 5$ to $\ell = 300$\footnote{The integration range for the iSW-effect as well as the details of instrumental noise and angular resolution are not very important, as most of the signal is at low $\ell$ below $\ell \sim 100$, due to the large cosmic variance provided by the primary CMB fluctuations, which is the largest source of noise.}, $\theta_\alpha$ are the cosmological parameters, ${\rm Cov}_{ij}(\ell)$ is the covariance of the spectrum $\bar C_{\zeta\gamma,i}(\ell)$ and is given by 
\be
{\rm Cov}_{ij}\left(\ell\right) = \frac{1}{2 \ell+1} \frac{1}{f_{\rm sky}} \left[\bar{C}_{\zeta\gamma,i}\bar{C}_{\zeta\gamma,j}(\ell)+\bar{C}_{\zeta\zeta}(\ell)\bar{C}_{\gamma\gamma,ij}(\ell) \right],
\ee
and where quantities with the bar represent the estimate of the signal, including intrinsic CMB fluctuations, instrumental noise and the beam of the CMB experiment as noise sources:
\bea
\bar{C}_{\zeta\gamma,i}(\ell) & = & C_{\zeta\gamma,i}(\ell) \label{eq:cross-noise}\\
\bar{C}_{\zeta\zeta}(\ell) & = & C_{\zeta\zeta}(\ell) + C_{\rm CMB}(\ell) + w_T^{-1}B^{-2}(\ell)\\
\bar{C}_{\gamma\gamma,ij}(\ell) & = & C_{\gamma\gamma,ij}(\ell)+\frac{\delta_{ij}}{n_i}
\eea
For Planck's noise levels, $w_T^{-1} = (0.02 \mu{\rm K})^2$ has been used and the beam was assumed to be Gaussian, $B^{-2}(\ell) = \left(2\times10^{-8}\right)^2\exp[\Delta\theta^2\ell\left(\ell+1\right) ]$,
with FWHM-width of $\Delta\theta = 7'.1$, corresponding to channels of Planck closest to the CMB-maximum at $\sim 160$ GHz. $n_i$ is the number of galaxies per steradian in the tomography bin $i$. We assume uncorrelated noise terms, and as a consequence the cross-spectra $C_{\zeta\gamma,i}(\ell)$ are unbiased estimates of the actual spectra, see Eq. (\ref{eq:cross-noise}). The spectrum $C_{\rm CMB}(\ell)$ of the CMB primary anisotropies from Planck has been computed with the CAMB code.

% ---  --- %
\subsection{Weak Lensing Fisher matrix}
The Fisher matrix for weak lensing is given by:
\be
F_{\alpha\beta}^{\rm WL} = 
f_{\rm sky}\sum_{\ell} \frac{\left(2\ell+1\right)}{2} 
\frac{\partial C_{\kappa\kappa,ij}(\ell) }{\partial \theta_\alpha}\bar{C}_{jk}^{-1}(\ell)
\frac{\partial C_{\kappa\kappa,km}(\ell) }{\partial \theta_\beta}\bar{C}_{mi}^{-1}(\ell)
\ee
where the sum runs from $\ell = 5$ to $\ell = 5000$, \citep[as from the official Euclid prescriptions, see][]{RedBook}, and where the sum over repeated indices is implied. We added a Poissonian shape noise term to the weak lensing spectra,
\be
\bar{C}_{\kappa\kappa,ij}(\ell) = 
C_{\kappa\kappa,ij}(\ell) + \delta_{ij}\frac{\langle \gamma_{\rm int}^{1/2}\rangle}{n_i},
\ee
$\gamma_{\rm int}$ is the rms intrinsic shear (here, we assume $\langle\gamma_{\rm int}^{1/2}\rangle$=0.22) 
and $n_{i}$ is the number of galaxies per steradians belonging to the $i$-th bin. We assume a Gaussian shape of the covariance while noting that non-Gaussian contribution can have a strong influence on the derived forecasts \citep{2009MNRAS.395.2065T}.

% ---  --- %
\subsection{Spectroscopic galaxy distribution Fisher matrix}
The galaxy  power spectrum Fisher matrix is given by \citep{seisen}
\begin{equation}
F_{\alpha\beta}^\mathrm{GC} = 
\int_{k_{\rm min}}^{k_{\rm max}}\frac{k^{2}{\rm d}k}{4\pi^{2}}
\frac{\partial\ln P_{\gamma\gamma}^\mathrm{spec}\left(z;k,\mu\right)}{\partial\theta_\alpha}
\frac{\partial\ln P_{\gamma\gamma}^\mathrm{spec}\left(z;k,\mu\right)}{\partial\theta_\beta}
\times V_{\rm eff},
\label{eq:FisherMatrix}
\end{equation}
where GC stays for galaxy clustering, the observed galaxy power spectrum $P_{\gamma\gamma}^\mathrm{spec}$ is given by Eq.~(\ref{eq:pk}), the derivatives are evaluated at the parameter values of the fiducial model, $k_{\rm min} = 0.001$ and $k_{\rm max}$ is such that the rms amplitude of the fluctuations at the corresponding scale $R_{\rm max} = 2 \pi/k_{\rm max}$ is $\sigma^2(R_{\rm max}) = 0.25$, with an additional cut at $k_{\rm max} = 0.2\, h/{\rm Mpc}$, in order to remain in the linear regime, 
 $V_{\rm eff}$ is the effective volume of the survey, given by
\begin{equation}
V_{\rm eff} \simeq   
\left(\frac{\bar n\,P_{\gamma\gamma}^\mathrm{spec}\left(k,\mu\right)}{\bar n\, P_{\gamma\gamma}^\mathrm{spec} \left(k,\mu\right)+1}\right)^{2}V_{\rm survey},
\label{eq:Volume}
\end{equation}
the latter equation holding for an average comoving number density $\bar n$.
 The number densities and further fiducial Euclid specifications can be found in \cite{RedBook}, \cite{Majerotto:2012mf}.

% ---  --- %
\subsection{Forecasts}
We computed forecasts on the measurement of $c_\mathrm{s}^2$ and $c_\mathrm{v}^2$ for a wide range of fiducial values, in order to capture the parameter determining capability of both experiments for a previously unknown set of parameters. The probes are assumed to be uncorrelated as discussed above, hence their Fisher-matrices add,
\begin{equation}
F_{\alpha\beta} = F_{\alpha\beta}^\mathrm{GC} + F_{\alpha\beta}^{\rm WL} + F^{\rm iSW}_{\alpha\beta},
\end{equation}
and we derive confidence contours on $c_\mathrm{s}^2$ and $c_\mathrm{v}^2$ and individual errors from this combined Fisher-matrix, marginalising over all other five parameters considered in this analysis. $F^\mathrm{GC}_{\alpha\beta}$ has been further marginalised over $P_\mathrm{shot}$, while the galaxy bias has been kept fixed.

\begin{figure}
\resizebox{\hsize}{!}{\includegraphics{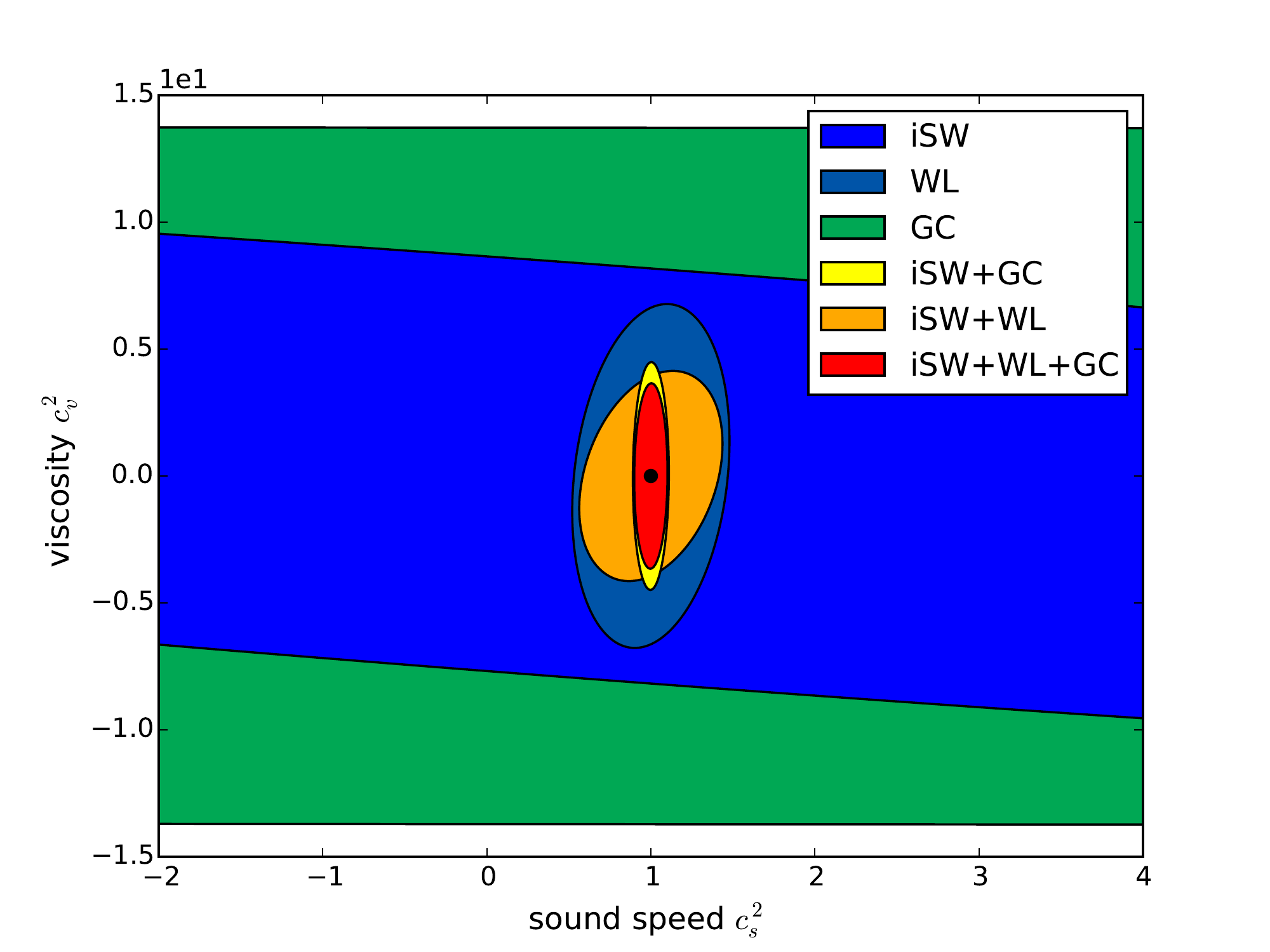}}
\caption{Forecasted $1\sigma$-constraints on $\cs$ and $\cv$ for individual probes and all possible combinations, for the fiducial choice $\cs=1$ and $\cv=0$.}
\label{fig:cs1cv0}
\end{figure}

\begin{figure}
\resizebox{\hsize}{!}{\includegraphics{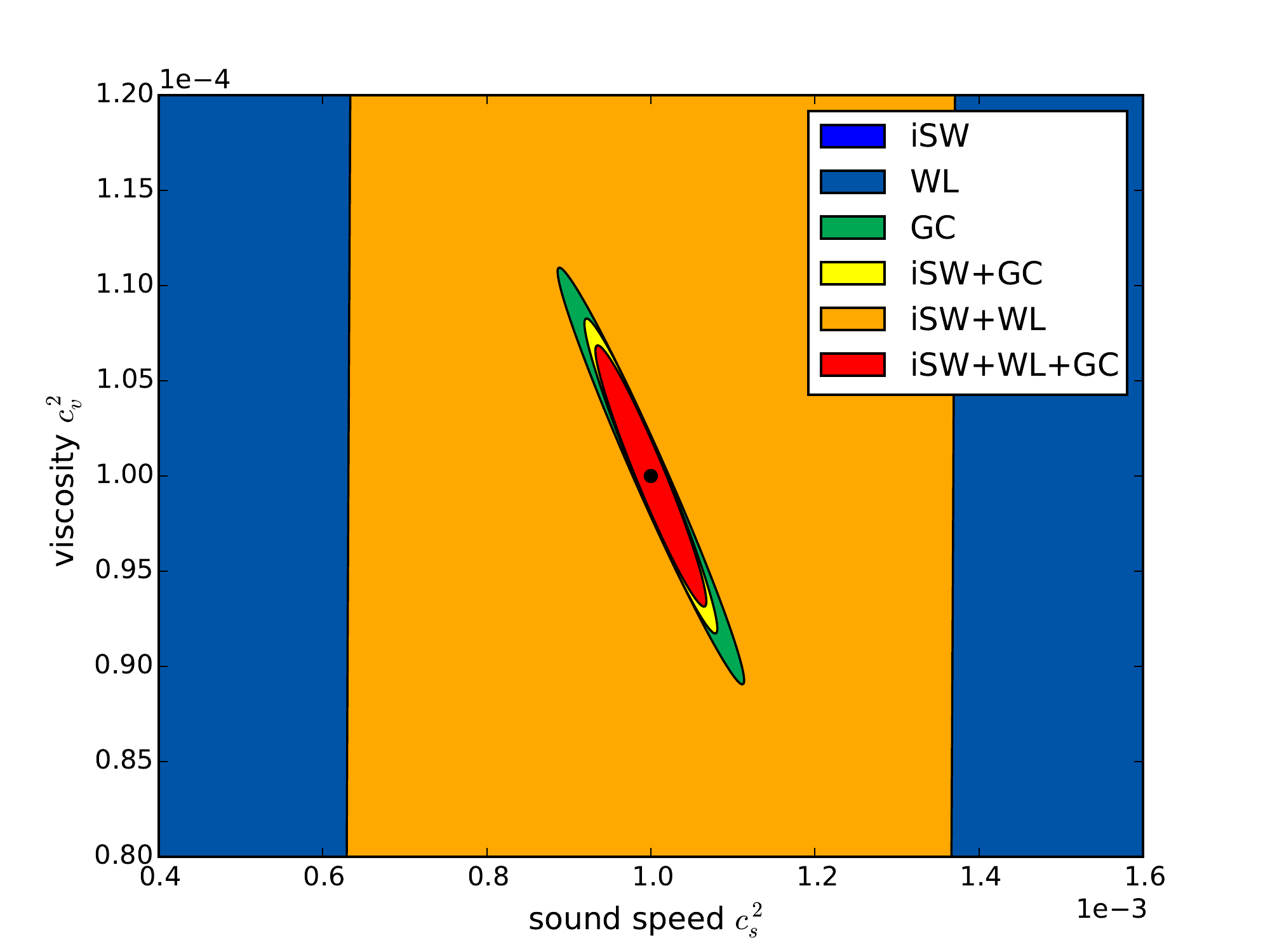}}
\caption{Forecasted $1\sigma$-constraints on $\cs$ and $\cv$ for individual probes and all possible combinations, for the fiducial choice $\cs=10^{-3}$ and $\cv=10^{-4}$.}
\label{fig:cs10e-3cv10e-4}
\end{figure}

\begin{figure}
\resizebox{\hsize}{!}{\includegraphics{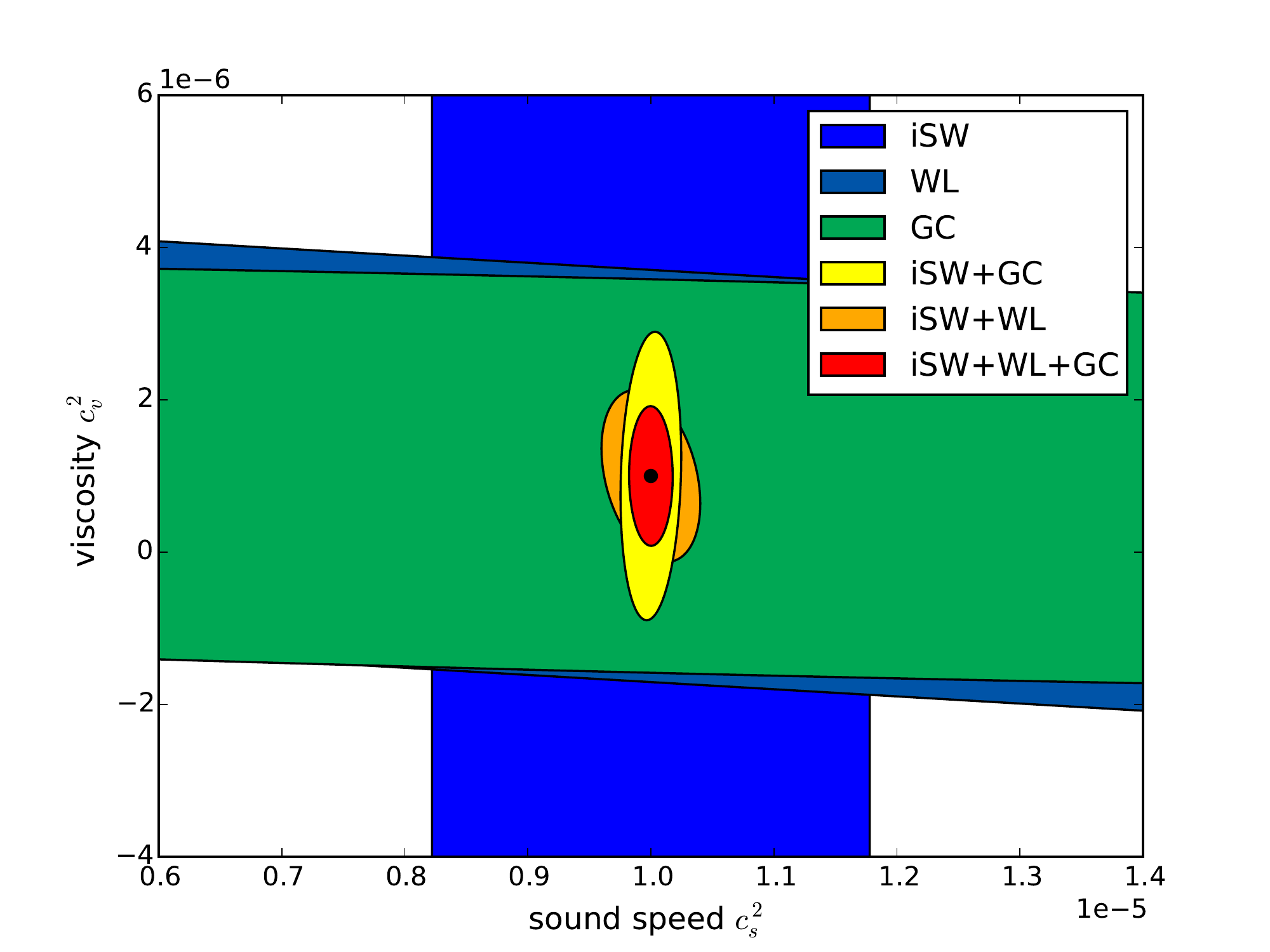}}
\caption{Forecasted $1\sigma$-constraints on $\cs$ and $\cv$ for individual probes and all possible combinations, for the fiducial choice $\cs=10^{-5}$ and $\cv=10^{-6}$.}
\label{fig:cs10e-5cv10e-6}
\end{figure}

\begin{figure}
\resizebox{\hsize}{!}{\includegraphics{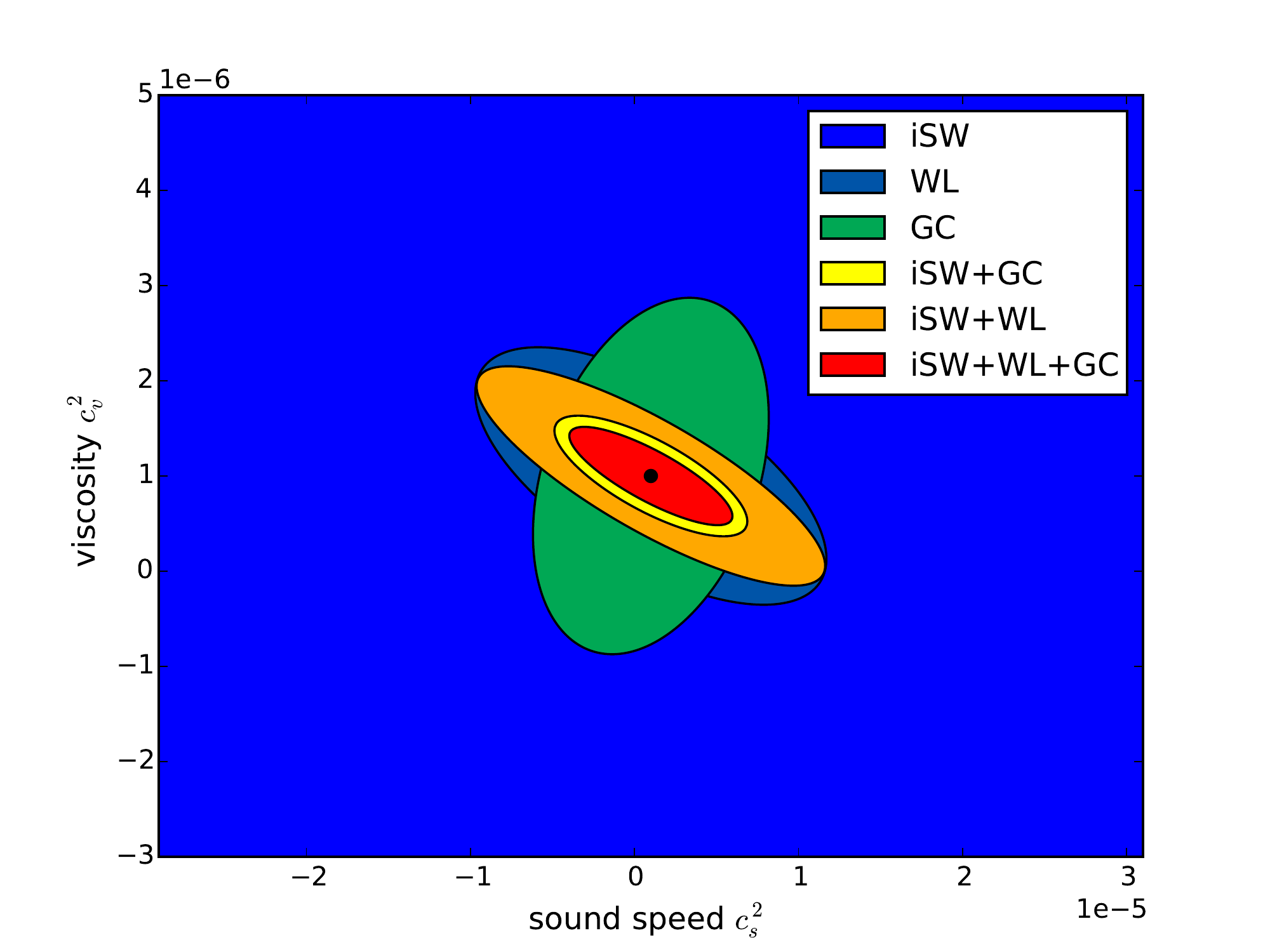}}
\caption{Forecasted $1\sigma$-constraints on $\cs$ and $\cv$ for individual probes and all possible combinations, for the fiducial choice $\cs=10^{-6}$ and $\cv=10^{-6}$.}
\label{fig:cs10e-6cv10e-6}
\end{figure}

Our forecasts on the following fiducial models:  $\{\cs, \cv  \} = \{1, \, 0\}$, $\{10^{-3}, \, 10^{-4}\}$, $\{10^{-5}, \, 10^{-6}\}$  and $\{10^{-6}, \, 10^{-6}\}$ are shown in Figs.~\ref{fig:cs1cv0}, \ref{fig:cs10e-3cv10e-4}, \ref{fig:cs10e-5cv10e-6} and~\ref{fig:cs10e-6cv10e-6}, respectively.
The first model corresponds to the case of simple scalar field dark energy, while the following two pairs of fiducial models were chosen such that $\cs = 10\, \cv$ because, as mentioned previously in Sec. \ref{subsect_pert}, the relevant quantity is the effective sound speed,  and the effect of $\cv$ in it is $\sim 10$ times stronger than that of $\cs$ when $w=-0.8$ because of the factor multiplying $\cv$ in Eq. (\ref{eq:ceff}). The last model, also having small $\cs$ and $\cv$, does not verify the latter relation, and has been chosen in order to be compared to previous work \citep{fingerprinting4} and to the similar case $\{\cs, \cv  \} =\{10^{-5}, \, 10^{-6}\}$.

In all plots, iSW constraints are shown in blue, weak lensing ones in dark blue, GC ones in green, combined iSW-GC ones in yellow, combined iSW-weak lensing ones in orange, and combined iSW-GC-weak lensing ones in red.

From Figs.~\ref{fig:cs1cv0}-\ref{fig:cs10e-6cv10e-6} it is clear that the results depend very much on the chosen fiducial model. A common feature is that iSW on its own does not provide very strong constraints. In particular (see also Tab.~\ref{tab:errors}),  $\cv$ is quite badly constrained, with relative errors ranging between $6.1 \times 10^4$ and $1.1 \times 10^5$, while relative errors on $\cs$ are much smaller: between $1.1 \times 10^{-1}$ and $1.4 \times 10$. This was to be expected since the iSW-effect along has a rather small signal strength of about $5\sigma$ for cross-correlating the CMB with the Euclid galaxy sample \citep{2008A&A...485..395D}.

\begin{table}
\begin{centering}
\begin{tabular}{|c|c|c|c|c|}
\hline 
\hline 
\hspace{0.5cm} &$\cs$\hspace{0.5cm}  & $\cv$  & $\sigma_{\cs}/\cs$ & $\sigma_{\cv}/\cv$ \tabularnewline
\hline 
\multirow{4}{*}{iSW}&$1$  & $0$  & $ 1.4 \times 10$  & $ \sigma_{\cv} = 8.6$ \tabularnewline
 
&$10^{-3}$  & $10^{-4}$  & $1.4 \times 10$ & $6.1 \times 10^{4} $ \tabularnewline

&$10^{-5}$  & $10^{-6}$  & $1.4 \times 10^{-1} $  & $1.1 \times 10^5$ \tabularnewline

&$10^{-6}$  & $10^{-6}$  & $4.0 \times 10^2 $  & $4.4 \times 10 $ \tabularnewline
\hline 
\multirow{4}{*}{WL}&$1$  & $0$  & $3.2 \times 10^{-1}$  & $ \sigma_{\cv} = 4.5$ \tabularnewline
 
&$10^{-3}$  & $10^{-4}$  & $1.6 \times 10^{2}$ & $1.4 \times 10^{2} $ \tabularnewline

&$10^{-5}$  & $10^{-6}$  & $3.7 \times 10 $  & $ 3.5 \times 10$ \tabularnewline

&$10^{-6}$  & $10^{-6}$  & $ 7.1$  & $8.9\times 10^{-1} $ \tabularnewline
\hline
\multirow{4}{*}{GC}&$1$  & $0$  & $ 1.1 \times 10^2$  & $ \sigma_{\cv} = 9.1$ \tabularnewline
 
&$10^{-3}$  & $10^{-4}$  & $ 7.5 \times 10^{-2}$ & $ 7.2 \times 10^{-2}$ \tabularnewline

&$10^{-5}$  & $10^{-6}$  & $ 2.2$  & $ 1.9$ \tabularnewline

&$10^{-6}$  & $10^{-6}$  & $ 4.7$  & $1.2 $ \tabularnewline
\hline
\multirow{4}{*}{iSW+WL}&$1$  & $0$  & $2.9 \times 10^{-1} $  & $ \sigma_{\cv} = 2.7$ \tabularnewline
 
&$10^{-3}$  & $10^{-4}$  & $ 2.5 \times 10^{-1}$ & $5.6 $ \tabularnewline

&$10^{-5}$  & $10^{-6}$  & $ 2.6 \times 10^{-2}$  & $7.5\times 10^{-1} $ \tabularnewline

&$10^{-6}$  & $10^{-6}$  & $ 7.0$  & $7.6 \times 10^{-1} $ \tabularnewline
\hline
\multirow{4}{*}{iSW+GC}&$1$  & $0$  & $7.2 \times 10^{-2} $  & $ \sigma_{\cv} = 3.0$ \tabularnewline
 
&$10^{-3}$  & $10^{-4}$  & $ 5.3 \times 10^{-2}$ & $ 5.5 \times 10^{-2} $ \tabularnewline

&$10^{-5}$  & $10^{-6}$  & $ 1.6 \times 10^{-2}$  & $1.2  $ \tabularnewline

&$10^{-6}$  & $10^{-6}$  & $3.9 $  & $4.2 \times 10^{-1} $ \tabularnewline
\hline
\multirow{4}{*}{GC+WL}&$1$  & $0$  & $6.7 \times 10^{-2}$  & $ \sigma_{\cv} = 3.7$ \tabularnewline
 
&$10^{-3}$  & $10^{-4}$  & $7.4 \times 10^{-2}$ & $7.1 \times10^{-2} $ \tabularnewline

&$10^{-5}$  & $10^{-6}$  & $1.9 $  & $1.7 $ \tabularnewline

&$10^{-6}$  & $10^{-6}$  & $3.5 $  & $4.8 \times10^{-1} $ \tabularnewline
\hline
\multirow{4}{*}{all}&$1$  & $0$  & $6.7 \times 10^{-2} $  & $ \sigma_{\cv} = 2.4$ \tabularnewline
 
&$10^{-3}$  & $10^{-4}$  & $4.5 \times 10^{-2}$ & $4.5 \times 10^{-2}$ \tabularnewline

&$10^{-5}$  & $10^{-6}$  & $1.2 \times 10^{-2} $  & $6.1 \times 10^{-1} $ \tabularnewline

&$10^{-6}$  & $10^{-6}$  & $3.3 $  & $3.4\times10^{-1} $ \tabularnewline
\hline
\end{tabular}
\par\end{centering}
\caption{Relative errors on the parameters $\cs$ and $\cv$ from iSW, weak lensing and GC alone, from the combination of iSW and WL, iSW and GC, GC and WL, and from all three datasets. For the case $\cv = 0$ the absolute error $\sigma_{\cv}$ is given.}
\label{tab:errors} 
\end{table}

Weak lensing constraints\footnote{With respect to \citet{fingerprinting4} we have improved the estimation of $P_\mathrm{NL}$ by using the full CAMB output instead of an analytical approximation to it.} are much stronger than iSW ones in the case of a fiducial scalar field dark energy, but become progressively comparable to them when the fiducial $\cs$ and $\cv$ become smaller, with the exception of the case $\cs = \cv = 10^{-6}$.

Even though both iSW and weak lensing do not give very strong constraints on sound speed and viscosity (see also Tab~\ref{tab:errors}), it is very interesting to notice that the two data sets complement each other very well. This is especially true for the case $\cs = 10^{-5}$ and $\cv = 10^{-6}$, represented in Fig.~\ref{fig:cs10e-5cv10e-6}, where the  blue ellipses, which indicate errors from iSW, have a very different degeneration direction with respect to the dark blue contours, corresponding to errors from weak lensing, but are comparable to them in size. Therefore the resulting combined errors are much smaller than those from a single dataset. In particular, the iSW effect gives better constraints on the sound speed and weak lensing on the viscosity parameter.  

Also, in the case of Figs.~\ref{fig:cs1cv0}, \ref{fig:cs10e-3cv10e-4} and \ref{fig:cs10e-6cv10e-6} iSW and weak lensing have different degeneracies, but here joining them does not improve the errors significantly because the weak lensing effect gives stronger constraints on both parameters. Here, the improvement in combining the two probes is rather the multiplication of a constraining likelihood with a wide one, resulting nevertheless in an increase in peakiness.

Also errors on $\cs$ and $\cv$ from GC are orthogonal to those from the iSW,  but only in the case where $\cs = 10^{-5}$ and $\cv = 10^{-6}$, see Fig. \ref{fig:cs10e-5cv10e-6}, this helps reducing the errors, because the former dataset performs better in constraining $\cv$ and the second $\cs$. In the case of fiducial $\cs = 10^{-3}$ and $\cv = 10^{-4}$ it is GC which gives best errors on both parameters, while for the $\cs = 1$ and $\cv = 0$ fiducial model, it is weak lensing.

Another interesting question is whether iSW adds important information to that provided by the other two datasets, which had already been analysed in \citet{fingerprinting4}. Table \ref{tab:errors} answers this question. It turns out that the information from iSW helps significantly in constraining $\cs$ and $\cv$ if the true model has $\cs = 10^{-5}$ and $\cv = 10^{-6}$, see Fig.~\ref{fig:cs10e-5cv10e-6}. In this case the iSW alone gives a strong constraint, which has moreover a different degeneration direction with respect to the error from galaxy clustering and weak lensing. For the other three fiducial models the gain when adding iSW is not very strong, as in both cases the combination of weak lensing and galaxy clustering gives a much tighter constraint in both the sound speed and the viscosity than the iSW alone.

It is interesting to notice (see Fig.~\ref{fig:cs10e-6cv10e-6}) that when both the sound speed and the viscosity are small, but the relation between $\cs$ and $\cv$ differs from $\cs \sim 10\cv$, the iSW effect error ellipse becomes much larger and as a result the sound speed parameter is less strongly constrained. Thus we conclude that $(i)$ very interesting results can be obtained through a combination from different cosmological probes and that $(ii)$ the iSW-effect is able to tighten constraints significantly for cases where there is a large difference between $\cs$ and $\cv$.

% --- section: summary --- %
\section{Summary}\label{sect_summary}
In this paper we have investigated how well the viscosity and sound speed of dark energy can be measured with the iSW cross-correlation spectrum, when using Planck and Euclid observations, and how joining iSW measurements to galaxy clustering and weak lensing ones improves constraints. 

We found that the speed of sound is quite well constrained, with relative errors as small as $0.14$ for small fiducial $\cs$ and $\cv$, while relative errors on the viscosity parameter are very large. Even though the anisotropic stress is not well constrained by the iSW, the error ellipses are interestingly orthogonal to those from weak lensing, hence the combination of these two datasets constrains tightly the parameter space, giving relative errors on $\cs$ and $\cv$ as small as $2.6 \times 10^{-2}$ and $7.5 \times 10^{-2}$ respectively. This is an improvement of a factor $\sim 1500$ in the measurement of the sound speed and $\sim 50$ in the measurement of the viscosity parameter, with respect to the weak lensing only constraint. The improvement obtained when combining iSW with galaxy clustering is smaller: a factor of $\sim 1.5$ in $\cv$ and $\sim 150$ in $\cs$. Finally, the addition of iSW to weak lensing and galaxy clustering constraints is most important if the fiducial sound speed and viscosity parameter are very small, while it is not very relevant for higher fiducial values of $\cs$.

It is also important to remind that in order to make the effect of dark energy perturbations stronger we have always used a value of the equation of state parameter $w=-0.8$. For values close to $w=-1$ the effects on the observables due to the dark energy perturbations are reduced, as all the phenomenological functions used (such as $Q(k,a)$) have a term $\propto (1+w)$. If we use a value of $w=-0.9$ we expect our final errors on the parameters to increase. But by how much? All the observables used in this paper depend most strongly on $Q^2$ (see Eq.~\ref{eq:qtot}) which is intrinsically included into the matter power spectrum; for a sound speed equal to zero $Q-1=(1+w)/(1-3w)a^{-3w}$ so the relative increase of the errors on the sound speed will be given by $1/[(Q(w=-0.9)-1)/(Q(w=-0.8)-1)]^2$ which is of about a factor $4$ larger, in agreement also with the results found in \citet{fingerprinting2}. 

A detection of sound speed and viscosity different from the values associated to a classical scalar field, i.e. $\cs = 1$ and $\cv = 0$, will point to a new understanding of the accelerated phase of the Universe. This is because the non ideal fluid considered in this paper can be thought of as en effective dark energy fluid parametrizing a modified gravity model, see \citet{Kunz:2006ca}. In practice, the detection of a zero sound speed does not automatically mean that we are dealing with an actual dark energy fluid, even though one would nevertheless experience effects which could be attributed to fluctuations of a fluid.

In this paper we found that joining data from Euclid and Planck we are able to constrain simultaneously the sound speed and the viscosity parameters, provided that the two are sufficiently small.  This is mostly due to the different sensitivity of the three observables, i.e. GC, WL and iSW to the two parameters. In most cases the iSW has a different degeneracy with respect to WL and GC, and this helps reducing the errors on $\cs$ and $\cv$ by a factor of $\sim 100$ (as pointed out before). Our results are in agreement with what found by \citet{Mota:2007sz, Calabrese:2010uf, Chang:2014mta}, who show that for values of $\cs$ approaching $\cs = 1$ the detection of a positive viscosity is very difficult, even when, as in the case of \citet{Calabrese:2010uf}, an early dark energy helps its detection by increasing its effect on smaller scales. 

To conclude, are Euclid and Planck able to measure the sound speed and the viscosity parameters of a dark energy component? If the values of $\cs$ and $\cv$ are small enough, the answer is yes; consequently, we will be able to constrain well the effective dark energy model. On the contrary, if sound speed and viscosity will escape detection, at least one of the two parameters will likely have large values. We would assume that other cosmological probes would not directly provide constraints on dark energy properties, but would nevertheless be able to provide constraining power by fixing other parts of the cosmological model, such as the dark matter density or the dark energy equation of state, which was not subject to variation in our investigation.

% --- section: acknowledgements --- %
\section*{Acknowledgements}
We acknowledge Martin Kunz and Luca Amendola for inspiring discussions.

E.~M. was supported by the Spanish MINECO's ``Centro de Excelencia Severo Ochoa"-programme under grant No. SEV-2012-0249 and by the Spanish MICINNs Juan de la Cierva programme (JCI-2010-08112), by CICYT through the project FPA-2012-31880, by the Madrid Regional Government (CAM) through the project HEPHACOS S2009/ESP-1473 under grant P-ESP-00346 and by the  European Union FP7 ITN INVISIBLES (Marie Curie Actions, PITN- GA-2011- 289442).
DS acknowledges financial support from the Fondecyt project number 11140496 and from the ``Anillo'' project ACT1122 founded by the ``Programa de Investigaci\'on asociativa''.

% --- section: bibliography --- %
\bibliography{bibtex/aamnem,bibtex/references}
\bibliographystyle{mn2e}

% --- section appendix --- %
\appendix
\bsp
\label{lastpage}
\end{document}